\documentclass{article}

\usepackage{arxiv}

\usepackage[utf8]{inputenc} 
\usepackage[T1]{fontenc}    
\usepackage{hyperref}       
\usepackage{url}            
\usepackage{booktabs}       
\usepackage{amsfonts}       
\usepackage{nicefrac}       
\usepackage{microtype}      
\usepackage{lipsum}
\usepackage{graphicx}

\usepackage{amsmath}
\usepackage{comment}
\usepackage{multirow}
\usepackage{makecell}
\usepackage{booktabs}
\usepackage{rotating}
\usepackage{dirtytalk}
\usepackage{tabularx}
\usepackage{xcolor}
\usepackage{hyperref}
\usepackage{array}
\usepackage{arydshln}
\usepackage{algpseudocode}
\usepackage{algorithm}

\title{A Privacy-Preserving Federated Learning Framework for Generalizable CBCT to Synthetic CT Translation in Head and Neck}

\author{ \href{https://orcid.org/0009-0008-2443-4029}{Ciro Benito Raggio\thanks{Corresponding author} \hspace{1mm}\includegraphics[scale=0.1]{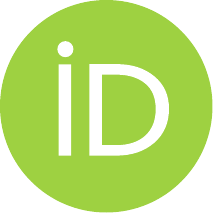}\hspace{1mm}}\\
  Institute of Biomedical Engineering\\
  Karlsruhe Institute of Technology\\
  Fritz-Haber-Weg 1, Karlsruhe 76131\\
  Baden-Württemberg, Germany \\
  \texttt{ciro.raggio@kit.edu} \\
\And
  \href{https://orcid.org/0000-0002-0219-0157}{Paolo Zaffino 
 \includegraphics[scale=0.1]{orcid.pdf}\hspace{1mm}} \\
  Department of Experimental and Clinical Medicine\\
  Magna Graecia University\\
  Viale Europa, Catanzaro 88100\\
  Italy
\And
  \href{https://orcid.org/0000-0002-5339-9583}{Maria Francesca Spadea
 \includegraphics[scale=0.1]{orcid.pdf}\hspace{1mm}} \\
  Institute of Biomedical Engineering\\
  Karlsruhe Institute of Technology\\
  Fritz-Haber-Weg 1, Karlsruhe 76131\\
  Baden-Württemberg, Germany
}

\begin{document}
\maketitle
\begin{abstract}
Cone-beam computed tomography (CBCT) has become a widely adopted modality for image-guided radiotherapy (IGRT). However, CBCT is characterized by increased noise, limited soft-tissue contrast, and artifacts. These issues result in unreliable Hounsfield unit (HU) values, which limits direct dose calculation. These issues were addressed by generating synthetic CT (sCT) from CBCT, particularly by adopting deep learning (DL) methods. However, existing DL approaches were hindered by institutional heterogeneity, scanner-dependent variations, and data privacy regulations that prevented multi-center data sharing.

To overcome these challenges, we proposed a cross-silo federated learning approach for CBCT-to-sCT synthesis in the head and neck region. This approach extended our original FedSynthCT framework to a different image modality and anatomical region. A conditional generative adversarial network (cGAN) was trained using data from three European medical centers within the SynthRAD2025 public challenge dataset while maintaining data privacy at each institution. 
A combination of the FedAvg and FedProx aggregation strategies, alongside a standardized preprocessing pipeline, was adopted to federate the DL model.

The federated model effectively generalized across participating centers, as evidenced by the mean absolute error (MAE) ranging from $64.38\pm13.63$ to $85.90\pm7.10$ HU, the structural similarity index (SSIM) ranging from $0.882\pm0.022$ to $0.922\pm0.039$, and the peak signal-to-noise ratio (PSNR) ranging from $32.86\pm0.94$ to $34.91\pm1.04$ dB. Notably, performance on an external validation dataset of 60 patients yielded comparable metrics: a MAE of $75.22\pm11.81$ HU, an SSIM of $0.904\pm0.034$ and a PSNR of $33.52\pm2.06$, confirming robust cross-center generalization despite differences in imaging protocols and scanner types, without additional training. Furthermore, a visual analysis of the results revealed that the obtained metrics were influenced by registration errors.

Our findings demonstrated the technical feasibility of FL for CBCT-to-sCT synthesis task while preserving data privacy, offering a collaborative solution for developing generalizable models across institutions without requiring data sharing or center-specific models.
\end{abstract}

\keywords{Federated Learning \and Synthetic Computed Tomography \and Deep Learning \and Image-To-Image Translation \and CBCT \and Head and Neck \and Data Sharing \and Data Privacy}

\section{Introduction}
\label{introduction}
Radiotherapy (RT) has seen significant advancements in recent decades, particularly with the integration of the image-guided radiotherapy (IGRT) technique. The IGRT involves the use of frequent imaging during treatment to account for anatomical changes and improve the accuracy of radiation delivery. Cone-beam computed tomography (CBCT) scanners are often incorporated into the gantry of linear accelerators, facilitating their integration into clinical practice. Consequently, CBCT emerged as one of the most widely adopted IGRT modalities~\cite{Boldrini_2023, Altalib2025, Spadea_Maspero2021}. 

The advent of CBCT addressed the limitations of planar imaging, which was the predominant modality in the early days of image guidance, by enabling volumetric imaging directly within the treatment suite. This development significantly impacted the precision of treatment and the streamlining of workflows~\cite{Boldrini_2023, Landry_Hua_2018, Rabe2025283}. Nevertheless, this approach shown improved alignment and visualization of patient anatomy, as well as real-time verification of target localization. Consequently, clinicians were able to manage both inter- and intra-fractional anatomical variability, and CBCT became a pivotal technology in the development of online image-guided adaptive radiotherapy (IGART) approaches~\cite{Landry_Hua_2018, Rabe2025283, RIOU2024603, Mastella2025}. 

However, CBCT is hindered by several drawbacks, including increased image noise, reduced soft-tissue contrast, and image reconstruction artifacts. These factors result in unreliable Hounsfield Unit (HU) values and limit its use in direct dose calculation~\cite{Spadea_Maspero2021, RIOU2024603, Rabe2025283}.  To address these challenges, the concept of synthetic CT (sCT) generation was introduced. In this context, CBCT images are translated into equivalent synthetic CT images with improved dosimetric fidelity, thereby avoiding superfluous radiation exposure and reducing the treatment workload by eliminating the need for additional scans~\cite{Altalib2025, Rabe2025283, HUIJBEN2024}.

Among the strategies developed for sCT generation, deep learning (DL) approaches have proven especially effective. Indeed, DL models demonstrated remarkable capabilities in producing high-quality synthetic images, suitable also for adaptive RT applications by learning complex, non-linear mappings between image domains~\cite{Spadea_Maspero2021, Altalib2025, Rabe2025283, HUIJBEN2024}. Despite the increasing adoption of sCT in real-world settings, its application is still limited by several challenges, which are rooted in institutional heterogeneity, scanner-dependent variations, and the usage of diverse imaging protocols. The aforementioned factors frequently necessitated single-site model training, a practice that often imposed constraints on clinical scalability~\cite{Altalib2025, Mastella2025, Raggio2025}. 

Although the aggregation of medical images from multiple sites appeared to be a straightforward solution to overcome the limitations of small or single-site datasets, this strategy is constrained by strict data privacy regulations, such as the Health Insurance Portability and Accountability Act (HIPAA)\footnote{\url{https://www.hhs.gov/hipaa/index.html}} in the United States and the General Data Protection Regulation (GDPR)\footnote{\url{https://gdpr-info.eu/}} in Europe. These legal frameworks impose restrictions on the sharing of sensitive health information, which made the centralization of imaging data across institutions a complex and often impracticable process~\cite{Raggio2025, kaissis2020secure, guan2024federated}.

These limitations emphasize the necessity for collaborative solutions that facilitate access to diverse and extensive datasets without necessitating direct data sharing or centralized storage.\\

To tackle these barriers, federated learning (FL) has gained attention as a decentralized training paradigm that enables institutions to collaboratively train DL models while keeping data local~\cite{Sandhu2023, DALMAZ2024103121, Hernandez-Cruz2024}. FL maintains compliance with data privacy regulations while concurrently enhancing cross-center collaboration without compromising security. The FedSynthCT-Brain framework was the first study to demonstrate the successful application of FL to the sCT synthesis task, specifically from brain T1-weighted magnetic resonance imaging (MRI) data, without compromising performance on external datasets~\cite{Raggio2025}. This finding suggested potential applications for extending the federated concept to the context of CBCT-to-sCT translation, where the development of robust, center-agnostic and generalizable models is essential for clinical adoption~\cite{HUIJBEN2024}.\\

In this study, we proposed a cross-silo horizontal FL approach for CBCT-to-sCT in the head and neck region, therefore extending the FedSynthCT~\cite{Raggio2025} framework to a different imaging modality. A federated DL model was collaboratively trained using data from different European university medical centers of the public SynthRAD2025 challenge dataset~\cite{thummerer2025synthrad2025}. The approach was validated not only on each federated center but also on an independent dataset outside the federation. This validation process was used to assess model generalization capabilities, thereby demonstrating that the proposed approach can result in more robust and generalizable models while preserving data privacy.

\section{Related work} \label{sec:related_works}
The CBCT-to-sCT conversion task has evolved considerably, with recent studies exploring its application across diverse anatomical regions, including the head and neck, thorax, pelvis, prostate, abdomen, and pancreas. This diversification underscores the extensive clinical pertinence and methodological challenges intrinsic to sCT generation. As anatomical areas have been the focus of increasing research over time, there has also been a growing interest in studying the DL models used, with the aim of increasing sCT accuracy~\cite{Rusanov2022, Altalib2025}.\\

Convolutional Neural Networks (CNNs) have been typically employed in the domain of medical image analysis. Their capacity to extract both local and global features renders them particularly well suited to enhancing CBCT image quality, a key necessity in the context of CBCT-to-sCT synthesis applications. Lately, the U-Net architecture~\cite{Ronneberger2015}, an encoder-decoder model based on CNNs, enabled pixel-level predictions while preserving the preservation of spatial and contextual information thanks to the skip connections, demonstrating notable efficacy in the image-to-image translation tasks~\cite{Rusanov2022, Altalib2025}.

Generative Adversarial Networks (GANs) have also been widely adopted for the CBCT-to-sCT task. Among them, the conditional GAN (cGAN) Pix2Pix~\cite{isola2018}, which was introduced specifically for image-to-image translation tasks, has gained particular popularity. This type of networks involved a generator-discriminator framework, wherein the generator synthesized realistic images often using an encoder-decoder structure, and the discriminator distinguished between real and synthetic data. Specifically, Pix2Pix employed a PatchGAN discriminator, which, unlike traditional CNN-based discriminators that output a single label for the entire image, was designed to assess image realism by locally analyzing small patches. This process penalized structural inconsistencies on a local scale and promoted the generation of accurate textures~\cite{isola2018}. 

Cycle-GANs built upon the foundation of the classic GAN concept by introducing cycle consistency, thereby ensuring the preservation of anatomical structures during the translation process between the CBCT and CT domains. Despite the encouraging results obtained with Cycle-GANs in enhancing sCTs for clinical application, several challenges emerged regarding training instabilities~\cite{Rusanov2022, Altalib2025}.\\

More recently, Transformer-based architectures and Denoising Diffusion Probabilistic Models (DDPMs) have been adopted to create increasingly realistic sCTs. Transformer-based architectures employed self-attention mechanisms to model long-range dependencies inherent in medical images. In contrast to CNNs, these models were capable of capturing broader spatial relationships. An interesting study that employed a Transformer-based architecture for this task was TransCBCT~\cite{Chen2022}, which adopted multi-head self-attention, positional encoding, and an encoder-decoder structure to effectively represent anatomical patterns in CBCT images. Nevertheless, it was observed that this category of models required extensive training data and substantial computational resources, while also being susceptible to image resolution~\cite{Chen2022, Altalib2025}.

On the other hand, DDPMs have been trained to reverse a noising process by gradually refining images from random noise. When conditioned on CBCT inputs, these models demonstrated to produce high-fidelity sCTs. The iterative denoising process enables the model to acquire knowledge regarding complex anatomical distributions. However, as mentioned for the transformer-based architectures, DDPMs shown to be computationally intensive. The process of inference has been observed to be slow, and it has been determined that a high level of hardware resources is necessary to employ them~\cite{Altalib2025}. Regardless of the architecture, each has demonstrated promising results in the accurate estimation of sCTs, including for radiotherapy planning~\cite{Rusanov2022, Altalib2025}. \\

The evaluation of sCT generation DL models in predicting accurate images comparable to gold standard planning CTs, also named ground-truth CT, was assessed by computing metrics such as mean absolute error (MAE), mean squared error (MSE), peak signal-to-noise ratio (PSNR), and structural similarity index (SSIM). Each metric was typically calculated within a specific region of interest, known as a mask, that allows elements outside the patient's body, such as the background, to be excluded from the evaluation.
Among these metrics, the MAE was often used as loss function, and was referred to as masked MAE or L1 Loss~\cite{Rusanov2022, Altalib2025}.

Gradient descent-based optimization has been extensively adopted for the iterative update of model weights. Concurrently, pre-processing methodologies such as image registration, image normalization and data augmentation have been demonstrated to be essential~\cite{Rusanov2022, LaGrecaSaintEsteven2023, HUIJBEN2024, Altalib2025}. \\

The application of federated learning (FL) in the medical imaging domain has seen growing interest in recent years, as evidenced by a surge of studies focusing on diagnostic and segmentation tasks across multiple imaging modalities~\cite{guan2024federated, Sheller2019-jo, Wenqi2019, Daiqing_Li2020, Chang2020}. FL offers a privacy-preserving framework that allows institutions to collaboratively train models while maintaining data locally, thus avoiding the need to share sensitive patient information and ensuring compliance with strict confidentiality requirements. Medical imaging datasets are often limited in size and subject to heterogeneity due to differences in acquisition devices, imaging protocols, and annotation standards. Moreover, data distributions across centers are typically non-independent and identically distributed (non-IID), which poses significant challenges for generalization. FL addressed these limitations by enabling distributed model training across diverse institutions and aggregating knowledge without transferring raw data, thereby enhancing robustness and reducing domain bias~\cite{Sandhu2023}.

Beyond the domains of classification and segmentation, federated approaches recently emerged in the context of medical image synthesis. For instance, FedMed-GAN has been developed to synthesize cross-modality brain MR images while leveraging simple federated aggregation techniques such as the federated averaging (FedAvg), which averaging the locally updated model weights from multiple clients to obtain a global model without sharing raw data~\cite{McMahan2017, Wang2023}. \\

Our previous work, named FedSynthCT-Brain was the first to demonstrate the feasibility of using FL for brain sCT generation from MRI, thereby establishing the foundation for sCT generation across institutions~\cite{Raggio2025}. The present study was built upon previous studies in the field and FedSynthCT-Brain, proposing an extension of the FL paradigm to a novel and clinically relevant task that has been addressed only through centralized settings, thus the sCT generation from CBCT images. 

\section{Materials and Methods}
\label{sec:met}
\subsection{Datasets}
\label{subsec:datasets}
The datasets employed in this work were from the publicly available SynthRAD2025 challenge repository~\cite{thummerer2025synthrad2025, SynthRAD2025DatasetV2}. The institutions (also referred to as centers or clients) focus on head and neck (HN) region and were labeled as centers A, B, C, and E. Center D was deliberately excluded from this study due to the limited availability. For consistency, we adopted the same naming convention for the centers as used in the challenge.

As reported in~\cite{thummerer2025synthrad2025}, a few thorax scans were included in the HN cohort; these outlier cases were excluded from the study. The final number of subjects included per center is reported in Table~\ref{tab:cbct_ct_parameters}. For each subject, the dataset included a CBCT scan, a CT scan, and a binary segmentation mask delineating the patient body.

\begin{table}[ht]
\centering
\small
\label{tab:cbct_ct_parameters}
\begin{tabularx}{\textwidth}{lXXXX}
\hline
\textbf{Parameter} & \textbf{Center A} & \textbf{Center B} & \textbf{Center C} & \textbf{Center E} \\
\hline
\multicolumn{5}{c}{\textbf{CBCT}} \\
\hline
Patient \# & 60 & 65 & 63 & 63 \\
Scanner & Elekta XVI v5.x & Elekta XVI v5.52 & Elekta XVI v5.x & Varian TrueBeam OBI \\
kVp & 100--120 & 100 & 120 & 100--125 \\
Exposure Time [ms] & 10--32 & 10 & 22 & 7500--18060 \\
Rows/Columns & 270 & 270 & \makecell[tl]{270--512 $\times$\\270--512} & 512 $\times$ 512 \\
Pixel spacing [mm] & 1 $\times$ 1 & 1 $\times$ 1 & 1 $\times$ 1 & \makecell[tl]{0.5--0.9 $\times$\\0.5--0.9} \\
Slice thickness [mm] & 1 & 1 & 1 & 2 \\
\hline
\multicolumn{5}{c}{\textbf{CT}} \\
\hline
Patient \# & 60 & 65 & 63 & 63 \\
Scanner & \makecell[tl]{Philips Big\\Bore (90),\\Siemens\\Biograph40\\ (10)} & Toshiba Aquilion/LB & \makecell[tl]{Philips\\Brilliance\\Big Bore (93),\\Siemens\\Biograph40 (7)} & \makecell[tl]{Toshiba\\Aquilion/LB,\\Siemens\\Biograph128} \\
kV & 120 & 120 & 120 & 120 \\
Exposure Time [ms] & 615--1000 & 500--1000 & 922--1457 & 500--1000 \\
Rows/Columns & 512 & 512 & 512 & 512 \\
Pixel spacing [mm] & \makecell[tl]{0.7--1.4 $\times$\\0.7--1.4} & \makecell[tl]{1.1--1.4 $\times$\\1.1--1.4} & \makecell[tl]{1--1.2 $\times$\\1--1.2} & \makecell[tl]{1--1.5 $\times$\\1--1.5} \\
Slice thickness [mm] & 2--3 & 1--3 & 2--3 & 3--5 \\
\hline
\end{tabularx}
\caption{Overview of the key CBCT and CT original acquisition parameters used at Centers A, B, C, and E according to~\cite{thummerer2025synthrad2025}}. 
\end{table}

Variability in imaging protocols and scanner types among relevant centers has been identified. Notably, CBCT scans from Centre E were acquired using a different system, in contrast to the devices employed in centres A, B and C (Varian TrueBeam OBI vs. Elekta XVI). Center E also exhibited markedly longer exposure times and higher in-plane resolution variability. Regarding the CT images, it was observed that all centers employed standard 120 kVp acquisitions; however, exposure time, slice thickness and pixel spacing varied.

\subsection{Federation setup and pre-processing}
\label{subsec:federation_setup}
The federated training setup included a central server associated with Center A dataset and three decentralized clients from Centers B, C, and E. The federated infrastracture was implemented using Flower~\cite{beutel2022flowerfriendlyfederatedlearning}, PyTorch~\cite{pytorch} and MONAI~\cite{cardoso2022monaiopensourceframeworkdeep}. Experiments were conducted on a single-node machine with an NVIDIA A100 (80GB), 16 CPUs, and 128GB RAM.\\

The Center A dataset was reserved exclusively for external validation, enabling an unbiased evaluation of the federated model's generalization performance. Within each participating client, four validation patients and four test patients were held out, while the remaining cases were used to train local models. Since data were released for a public challenge, several pre-processing operations were applied by the authors prior to the publication, such as isotropic resampling, defacing, and rigid registration between modalities~\cite{thummerer2025synthrad2025}. We performed an additional deformable registration process on the test data from each center to minimize the impact of registration on the model evaluation.\\

To ensure consistent spatial resolution across the federation, each client applied a standardized pre-processing pipeline to locally adapt data to a $256 \times 256 \times 256$ dimension. This decentralized approach --which avoided any inter-client data exchange-- involved conditional cropping followed by resizing and constant-padding as needed. Furthermore, an intensity processing was applied. A local normalization process was performed on each client for each CBCT scan, employing a LUT-based normalization method proposed by Vicario et al.~\cite{Vicario2022} with a fixed intensity range of [-800, 2000]. Subsequently, CT scans were clipped to the range [-1000, 1500] to mitigate the impact of extreme values introduced by metal implants. These ranges were empirically chosen by analyzing the CBCTs and CTs intensity distributions available at each client site, without exchanging any images. 

This preliminary harmonization step allowed for the approximation of a common intensity range across clients while preserving the constraints of the federated paradigm.

\subsection{Deep learning model}

The DL architecture employed relied on the cGAN paradigm, and more specifically, on the original Pix2Pix architecture~\footnote{https://github.com/phillipi/pix2pix}~\cite{isola2018}. Pix2Pix has been identified as one of the most widely adopted architectures for the CBCT-to-sCT translation task, as indicated in Section~\ref{sec:related_works}. The adversarial paradigm exhibited a remarkable capacity in managing imperfectly paired or partially aligned data, a prevalent scenario in clinical practice where precise voxel-wise correspondence between images cannot be assured, even after the registration process. This has led to the identification of Pix2Pix as a suitable and practical solution for our application setting.\\

The generator was structured as a 2D U-Net with eight encoder-decoder blocks connected via skip connections, and included dropout layers to mitigate overfitting. The discriminator was implemented as a 2D PatchGAN, composed of convolutional blocks combining 2D convolutions, instance normalization, and LeakyReLU activations to assess local realism at the patch level.\\

The training of the DL model was implemented using the Random-Multi2D sampling approach, to enhance the robustness and generalisability. This approach consisted of presenting to the model different axial, sagittal, and coronal slices in a randomized order, ensuring that the DL model received non-sequential input from multiple anatomical planes~\cite{Raggio2025}. \\

In detail, the discriminator's objective was achieved by minimizing the sum of two sigmoid cross-entropy losses:
\begin{equation}
    \mathcal{L}_{\text{D}} = \frac{1}{2} \times [\mathrm{BCE}(\mathrm{\mathcal{D}}(CT, CBCT), \mathbf{1}) + \mathrm{BCE}(\mathrm{\mathcal{D}}({sCT, CBCT}), \mathbf{0})]
\end{equation}
where $\mathrm{BCE}(\cdot)$ was the cross-entropy loss output, $\mathrm{\mathcal{D}}(\cdot)$ was the discriminator output, and 1 and 0 were images of ones and zeros indicating real and fake labels, respectively. The first term penalized misclassifications of CTs as fake, while the second term penalized misclassifications of sCTs as real.

The adversarial loss function employed for the optimization of the generator was combined by two components. 
The first was a sigmoid cross-entropy loss that enabled the discriminator to classify the generated images as real or fake: 

\begin{equation}
    \mathcal{L}_{\text{GAN}} = \mathrm{BCE}(\mathrm{\mathcal{D}}(sCT, CBCT), \mathbf{1}).
\end{equation}

The second term was the L1 loss, specifically the Masked MAE (see Equation \ref{eq:masked_mae}) --typically employed for this task as reported in Section \ref{sec:related_works}-- between the generated sCT and the ground-truth CT, which enforced structural similarity and penalized large deviations. This loss was also referred to as pixel loss:

\begin{equation}
    \mathcal{L}_{\text{pixel}} = \mathrm{MAE}(CT - {sCT}).
\end{equation}

The total generator loss was subsequently computed as:
\begin{equation}
    \mathcal{L}_{\text{G}} = \mathcal{L}_{\text{GAN}} + \lambda_{pixel} \cdot \mathcal{L}_{\text{pixel}},
\end{equation}
where $\lambda_{pixel} = 100$ as reported in~\cite{isola2018}.\\

Inference was based on the median voxel-voting mechanism. Specifically, following the generation of the slice-wise predictions along each anatomical plane, the final sCT volume was computed by taking the median across the three orientations~\cite{Spadea_Maspero2021, Raggio2025}. Both training and inference strategies were previously validated in the federated scenario within the FedSynthCT-Brain framework~\cite{Raggio2025}. \\

In each local client, the model was trained over 2 local epochs using a batch size of 32 slices before participating in a federated aggregation round. The Adam optimizer with a learning rate of $10^{-4}$ was used. 

To further improve robustness and cross-site generalization, a consistent data augmentation pipeline was applied at the client level. Each client independently introduced random geometric and intensity transforms during the training phase, including flipping, rotations, translations, and random rescaling or shifting of intensity values.

\subsection{Aggregation strategy}

The combination of FedAvg and FedProx~\cite{li2020federatedepochsopt}, was employed as aggregation strategy for this study. The aggregation was equivalent with the approach used in the FedSynthCT-Brain study, as it was identified as the optimal aggregation strategy, yielding results that are equivalent to results obtained by alternative strategies while requiring a reduced number of rounds~\cite{Raggio2025}.

The FedAvg algorithm was implemented on the server side to aggregate the clients' weights, ensuring that the contribution of each client was proportional to the size of its dataset. Therefore, in the Equation \ref{eq:fedavg}, \( K \) represents the total amount of clients; \( n_k \) denotes the amount of samples for a single client \( k \); \( n \) is the total amount of samples across the federation, and \( w_{t+1}^k \) represents the post-training weights of the client \( k \). 

\begin{equation} 
\label{eq:fedavg}
w_{t+1} = \sum_{k=1}^{K} \frac{n_{k}}{n}w_{t+1}^k
\end{equation}

The FedProx method was subsequently implemented on the client side, thereby introducing a proximal term, denoted by \( \frac{\mu}{2}||w-w_{t}||^2 \) to the local objective function (thus, the Masked MAE loss). This term penalized large deviations from the global model weights in order to control local update variations. The Equation \ref{eq:proximalterm} provides a formal explanation of the impact of the proximal term on the client model. The original local objective is denoted by  \( F_{k}(w) \), the pre-training global model weight by \( w_t \), the weights during local training by \( w \),  and the proximal coefficient by \( \mu \). Consequently, FedProx mitigates update divergence and approximates FedAvg when \( \mu \) = 0, as evidenced in~\cite{li2020federatedepochsopt}.

\begin{equation} \label{eq:proximalterm}
h_{k}(w,w_{t}) = F_{k}(w) + \frac{\mu}{2}||w-w_{t}||^2
\end{equation}

In the proposed work, the \( \mu \) term was set to 3, in accordance with the findings of the FedSynthCT-Brain study~\cite{Raggio2025}.

\subsection{Evaluation}

The performance of the federated model was evaluated using standard image similarity metrics on the external test dataset (Center A) and the client test datasets (Centers B, C, and D)~\cite{Spadea_Maspero2021, Rusanov2022, Altalib2025}. Specifically, the following metrics were employed:

\begin{itemize}
    \item \textbf{Masked Mean Absolute Error}:
    \begin{equation}
    \label{eq:masked_mae}
        \text{MAE} = \frac{1}{n} \sum_{i=1}^{n} \left| CT_i - sCT_i \right|,
    \end{equation}
    where $n$ is the number of voxels within the region of interest (ROI).
    
    \item \textbf{Masked Structural Similarity Index Measure (SSIM)}:
    \begin{equation}
        \text{SSIM} = \frac{(2 \mu_{\text{sCT}} \mu_{\text{CT}} + C_1)(2 \sigma_{\text{sCT,CT}} + C_2)}{(\mu_{\text{sCT}}^2 + \mu_{\text{CT}}^2 + C_1)(\sigma_{\text{sCT}}^2 + \sigma_{\text{CT}}^2 + C_2)},
    \end{equation}
    where the mean ($\mu$), variance/covariance ($\sigma$), and constants $C_1 = (k_1 L)^2$, $C_2 = (k_2 L)^2$ are computed over the ROI. Here, $L$ is the dynamic range of the CT image, with $k_1 = 0.01$ and $k_2 = 0.03$.
    
    \item \textbf{Masked Peak Signal-to-Noise Ratio (PSNR)}:
    \begin{equation}
    \text{PSNR} = 10 \cdot \log_{10} \left( \frac{MAX_{CT}^2}{\text{MSE}} \right),
    \end{equation}
    where $MAX_{CT}$ is the maximum intensity value in the CT image and MSE is the mean squared error between CT and sCT within the ROI.
\end{itemize}

\section{Results and Discussions}
\label{sec:res}

\begin{figure}[H]
    \centering
    \includegraphics[width=1\linewidth]{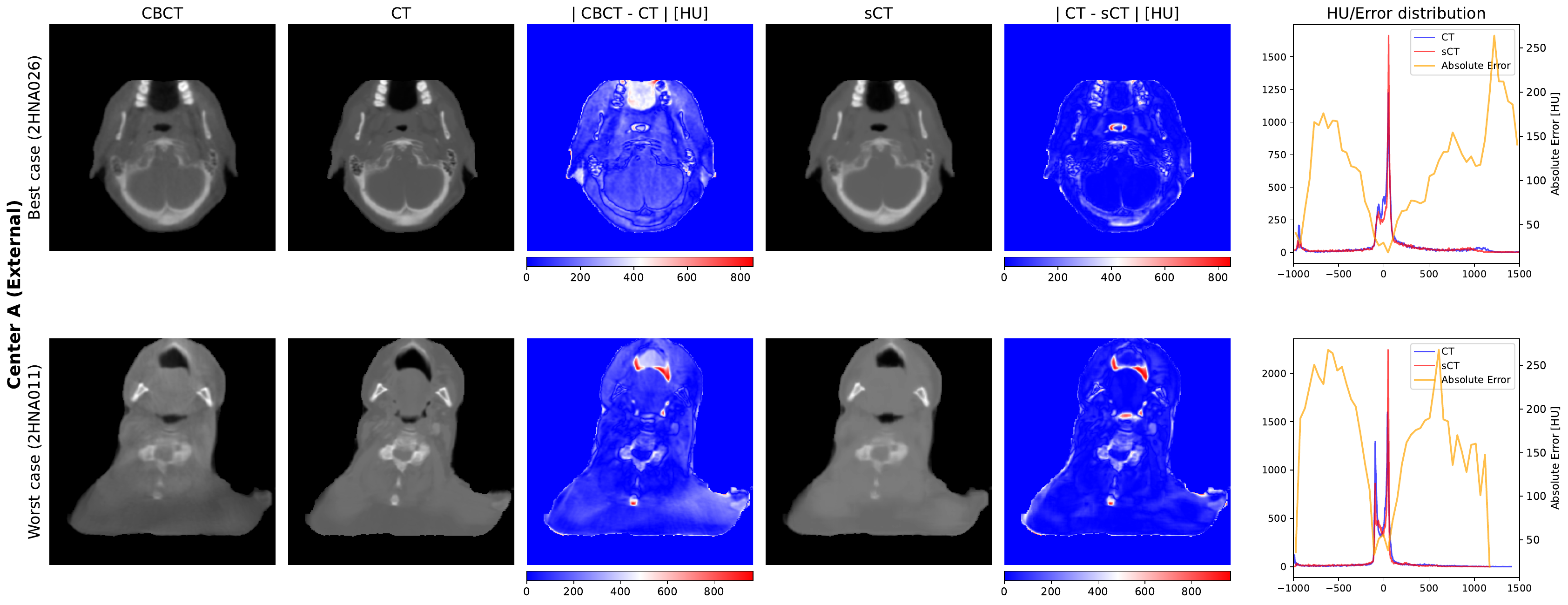}
    \caption{Visual evaluation of the best and worst sCT cases from Center A, used exclusively as an external test set. For each case, the mid-axial slice of the original CBCT, the ground-truth CT, the generated sCT, and the absolute difference maps for CBCT-to-CT and sCT-to-CT (error map) are shown. Additionally, intensity distributions of the sCT and CT for the displayed slice are presented, overlaid with the pixel-wise MAE. These distributions highlight the conversion accuracy across different tissue density ranges. Notably, regions with very high absolute differences appear in both CBCT-to-CT and error maps, suggesting the presence of residual misregistration between CBCT and CT scans.}
    \label{fig:center_a_results}
\end{figure}

\begin{figure}[ht]
    \centering
    \includegraphics[width=1\linewidth]{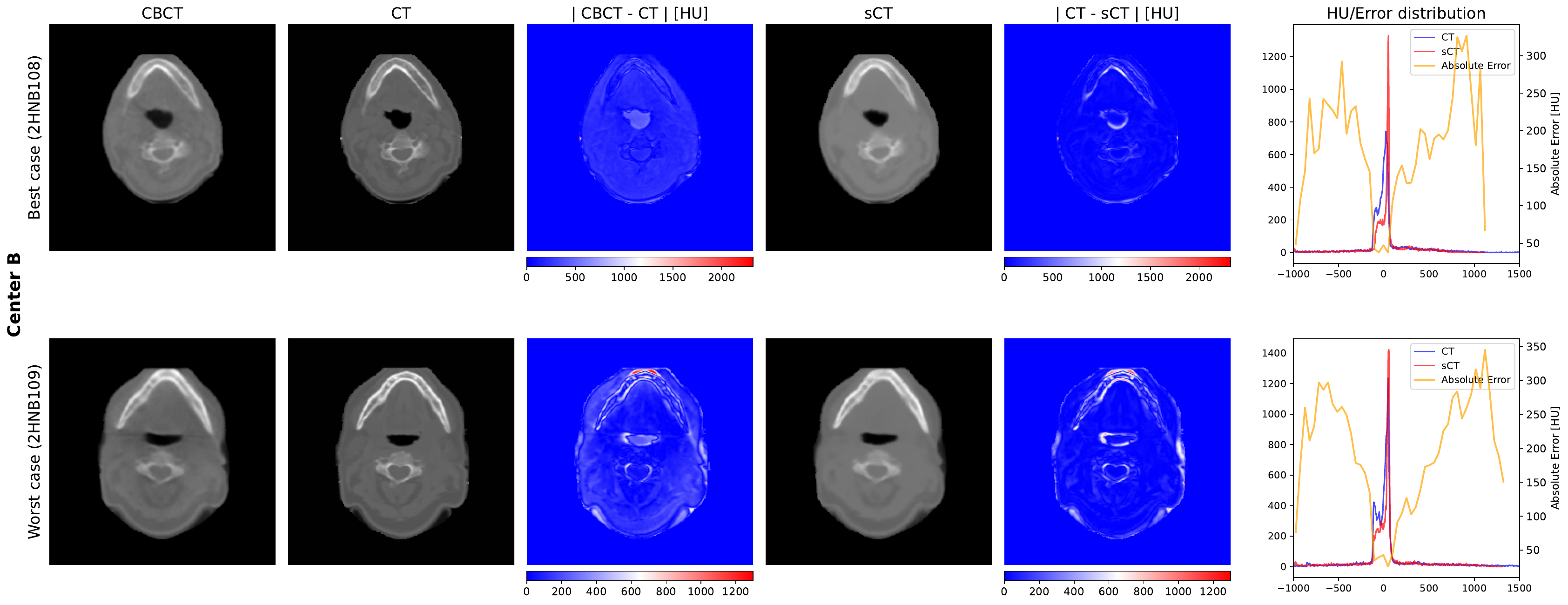}
    \caption{Qualitative results for the best and worst sCT cases from Center B, which showed lower image similarity metrics among the training centers. Widespread registration artifacts between CBCT and CT scans contributed to the visible errors. This was especially evident in the oral cavity of the 2HNB109 patient, and are reflected in both $|CBCT-CT|$ figure and error map.}

    \label{fig:center_b_results}
\end{figure}

\begin{figure}[H]
    \centering
    \includegraphics[width=1\linewidth]{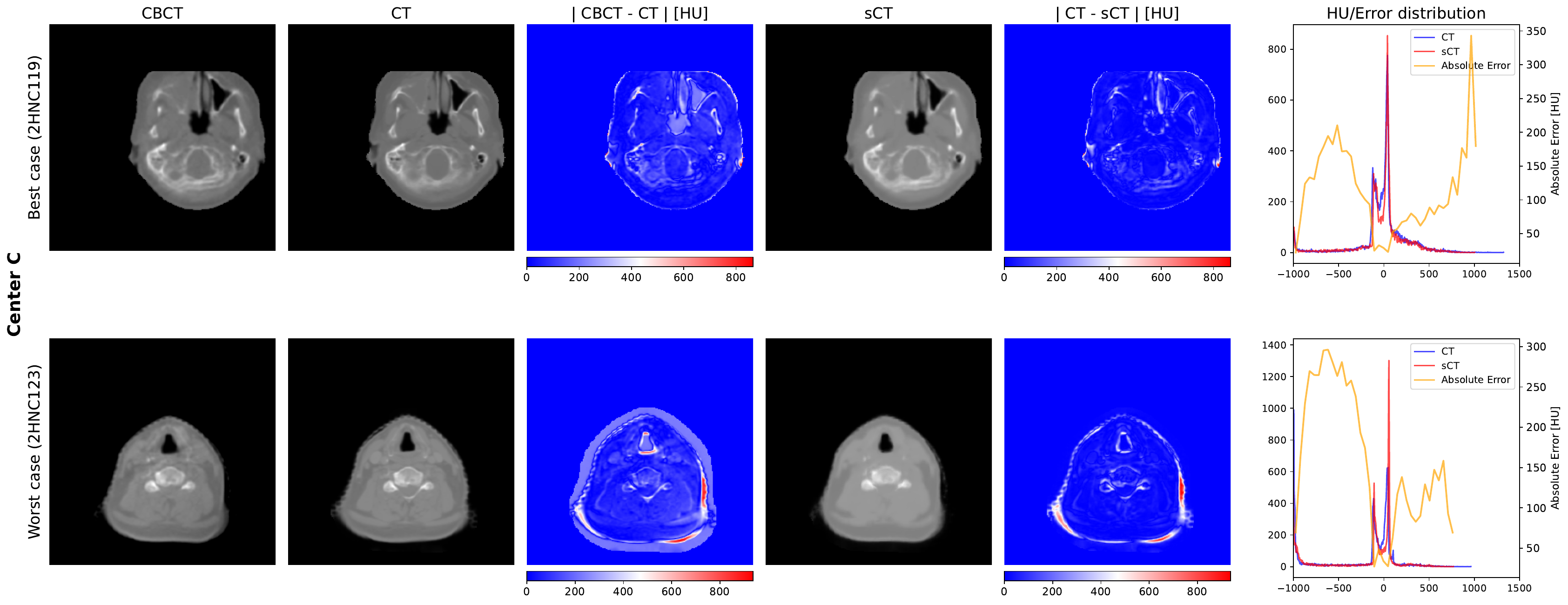}
    \caption{Visual comparison of the best and worst sCT cases from Center C. Compared to the other centres, Center C showed fewer registration-related errors and better image similarity, as evidenced by both the error maps ($|CT - sCT|$) and the distribution plots. However, case 2HNC123, identified as the worst-performing example, exhibited evident misalignments prior to the sCT generation despite the additional registration process.}
    \label{fig:center_c_results}
\end{figure}

\begin{figure}[H]
    \centering
    \includegraphics[width=1\linewidth]{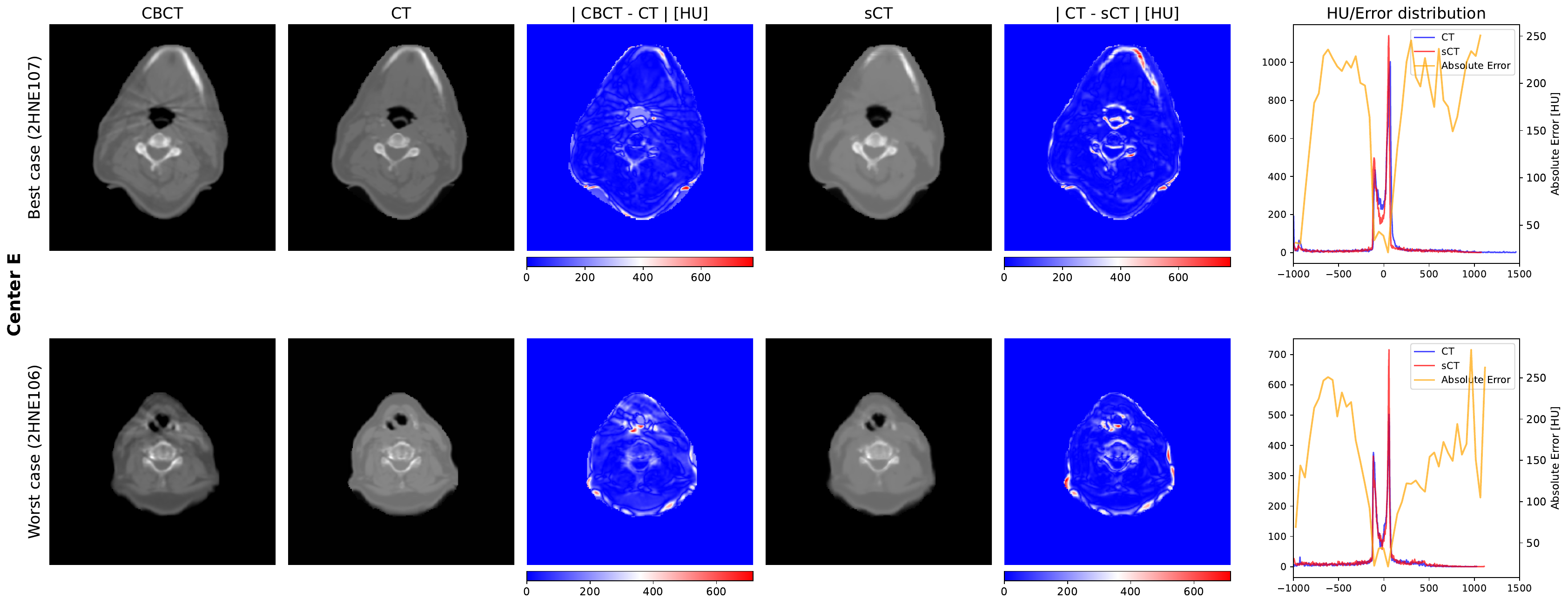}
    \caption{Visual evaluation of the best and worst sCT cases from Center E. Despite the use of a different scanner and acquisition protocol, the model maintained strong performance, with distribution plots indicating consistent HU alignment and localized discrepancies, mostly in air and bone regions due to inter-scan misalignments.}

    \label{fig:center_e_results}
\end{figure}

The performance of the proposed federated cGAN model across multiple clinical centers was assessed through both quantitative and qualitative results, as reported in Table~\ref{tab:results_by_center} and Figures~\ref{fig:center_a_results}--\ref{fig:center_e_results}, respectively.

\begin{table}[h]
\centering
\begin{tabular}{lcccc}
\toprule
\textbf{Center} & \textbf{MAE [HU]} & \textbf{SSIM} & \textbf{PSNR [dB]}\\
\midrule
Center A (external) & 75.22 $\pm$ 11.81 & 0.904 $\pm$ 0.034 & 33.52 $\pm$ 2.06 \\
Center B & 85.90 $\pm$ 7.10  & 0.882 $\pm$ 0.022 & 32.86 $\pm$ 0.94 \\
Center C & 64.38 $\pm$ 13.63 & 0.922 $\pm$ 0.039 & 34.49 $\pm$ 1.39 \\
Center E & 74.22 $\pm$ 13.23 & 0.915 $\pm$ 0.013 & 34.91 $\pm$ 1.04 \\
\bottomrule
\end{tabular}
\caption{Image similarity metrics obtained by the federated model on each client. The results are reported as mean $\pm$ standard deviation for mean absolute error (MAE), structural similarity index (SSIM), and peak signal-to-noise ratio (PSNR). Center A was used exclusively for external testing, while Centers B, C, and E participated in the federated training.}
\label{tab:results_by_center}
\end{table}

As detailed in Section~\ref{subsec:federation_setup}, Center A dataset was excluded from the federated training process and served as an external test set, including 60 test cases which provided a more robust and less biased estimation of the model’s generalization capabilities. Centers B, C, and E contributed data to train the federated model, with each center reserving four patients for the local evaluation.

As shown in Table~\ref{tab:results_by_center}, despite not participating in the federated training, Center A achieved performance metrics comparable to those of the training centers, with a MAE of $75.22 \pm 11.81$ HU, an SSIM of $0.904 \pm 0.034$, and a PSNR of $33.52 \pm 2.06$ dB. 

A qualitative evaluation of the best and worst test cases from Center A (Figure \ref{fig:center_a_results}) revealed that sCTs aligned closely with the ground-truth CT. The error maps --obtained as $|CT - sCT|$-- showed that most deviations were within $\pm100$ HU, demonstrating the model’s ability in the sCT generation task. However, the absolute difference between the original CBCT and the ground-truth CT, also presented in Figure~\ref{fig:center_a_results} as $| CBCT - CT |$, revealed spatial discrepancies primarily due to registration errors and anatomical variations that occurred during treatment. 

For instance, anatomical differences in the oral cavity were observed in patient 2HNA011 (Figure \ref{fig:center_a_results}). These errors were inherited by the sCT, thereby degrading its quality for the evaluation process and negatively affecting the reported metrics (Table~\ref{tab:results_by_center}).

Despite these registration challenges, the overall evaluation confirmed that the federated model generalized effectively to unseen data without requiring additional training or fine-tuning.\\

Among the federated clients, Center B obtained a $\text{MAE} = 85.90 \pm 7.10$ HU, an $\text{SSIM} = 0.882 \pm 0.022$, and a $\text{PSNR} = 32.86 \pm 0.94$ dB, reflecting lower performance in image similarity compared to Centers C, E, and the external Center A. Nevertheless, qualitative results (Figure~\ref{fig:center_b_results}) demonstrated that the sCTs were anatomically consistent with the ground-truth CTs, particularly in the representation of major structures. As observed for Center A, misalignments between the input CBCT and ground-truth CT were found to be the primary source of residual errors in the generated sCT. A detailed analysis confirmed that Center B exhibited a higher prevalence of registration issues. These findings provided a rationale for explaining the observed discrepancy in the performance metrics of Center B, which exhibited lower performance on average compared to the model's performance in the other centers.

Center C yielded the best overall results in terms of image similarity metrics, with the lowest MAE ($64.38 \pm 13.63$ HU), highest SSIM ($0.922 \pm 0.039$), and highest PSNR ($34.49 \pm 1.39$ dB). Notably, Center C exhibited a reduced influence from registration errors when compared to the other centers, thus obtaining better image similarity metrics. In the worst case scenario (2HNC123) presented in Figure \ref{fig:center_c_results}, the same phenomenon as in previous analyses was observed: major errors were related to misalignments of the original CBCT and the ground-truth CT, rather than pure generation errors.

Interestingly, Center E achieved results that were well-aligned with the federation, despite images being acquired with a completely different scanner and acquisition protocol (see Table~\ref{tab:cbct_ct_parameters}). The federated model achieved a $\text{MAE} = 74.22 \pm 13.23$ HU, an $\text{SSIM} = 0.915 \pm 0.013$, and a $\text{PSNR} = 34.91 \pm 1.04$ dB (Table~\ref{tab:results_by_center}). Furthermore, quantitative results confirmed the quality of the generated sCTs, as shown in Figure \ref{fig:center_e_results}.\\

A subsequent analysis of the results, with a focus on anatomical structures and HU distributions across all centers, demonstrated the model's ability to preserve tissue-specific intensity patterns. The most substantial errors were consistently associated with regions containing tissues that have a smaller overall volume in the head and neck area, such as air pockets and dense bone. These tissues were underrepresented in the training data relative to soft tissues, making them more challenging to predict accurately. Furthermore, due to the frequent misalignment of test data, errors in these regions had a considerable impact on image similarity metrics, amplified by their high HU values. As no objective and independent method is available to quantify registration errors, they could only be assessed qualitatively through visual inspection of the results.

\section{Conclusion}
\label{sec:conc}
This study demonstrated the feasibility and effectiveness of a cross-silo FL approach for CBCT-to-sCT synthesis in the head and neck region, using the publicly available SynthRAD2025 challenge dataset~\cite{thummerer2025synthrad2025}. It extended the previously developed FedSynthCT-Brain framework~\cite{Raggio2025}, originally designed for brain MRI-to-sCT synthesis by leveraging FL as a privacy-preserving and collaborative solution.\\

The findings demonstrated that the federated model generalized effectively across the participating centers and on an external validation dataset, without requiring additional training or fine-tuning, thereby maintaining consistent performance despite differences in imaging protocols, scanner types, and acquisition parameters. This mitigates the impact of institutional heterogeneity, a common limitation in single-center model deployments.

These results align with requirements outlined in recent related work, such as the SynthRAD2023 challenge report~\cite{HUIJBEN2024}, which emphasized that a model should ideally be applicable across different centers without the need for conditional tuning. This is especially relevant given that current commercial solutions are not tailored to individual institutions.\\

Although an additional deformable registration step was required for the evaluation data, the observed performance differences across centers were primarily attributed to image misalignments. Qualitative analysis confirmed that the generated sCTs were anatomically consistent with the corresponding ground-truth CTs. However, the model exhibited some limitations, with the most prominent errors occurring in regions with extreme HU values --such as air pockets and dense bone-- that were underrepresented in the training data when using a whole-slice training strategy. 

These findings suggest potential future directions, including more targeted training strategies, such as the multi-discriminator framework proposed by Kumar et al.~\cite{Kumar2024}, which aims to improve generator performance by better capturing anatomically underrepresented structures. While the clinical integration of FL-generated sCTs into RT treatment planning represents a potential application of this work, its evaluation was not possible in the present study due to the absence of treatment planning data in the SynthRAD2025 dataset, which contains only imaging data. Future research may focus on evaluating the clinical applicability of sCTs generated through FL.

\section*{Data availability}
The data used for the study were extracted from the public SynthRAD2025 Grand Challenge dataset and are available at \url{https://zenodo.org/doi/10.5281/zenodo.15373853}.

\bibliographystyle{unsrt}

\begin{thebibliography}{10}

\bibitem{Boldrini_2023}
Luca Boldrini, Andrea D’Aviero, Francesca De~Felice, Isacco Desideri, Roberta Grassi, Carlo Greco, Giuseppe~Carlo Iorio, Valerio Nardone, Antonio Piras, and Viola Salvestrini.
\newblock Artificial intelligence applied to image-guided radiation therapy {(IGRT)}: a systematic review by the young group of the italian association of radiotherapy and clinical oncology (yairo).
\newblock {\em La Radiologia Medica}, 129(1):133–151, September 2023.

\bibitem{Altalib2025}
Alzahra Altalib, Scott McGregor, Chunhui Li, and Alessandro Perelli.
\newblock Synthetic {CT Image Generation From CBCT}: A systematic review.
\newblock {\em IEEE Transactions on Radiation and Plasma Medical Sciences}, pages 1--1, 2025.

\bibitem{Spadea_Maspero2021}
Maria~Francesca Spadea, Matteo Maspero, Paolo Zaffino, and Joao Seco.
\newblock Deep learning based {synthetic‐CT} generation in radiotherapy and {PET}: A review.
\newblock {\em Medical Physics}, 48(11):6537–6566, August 2021.

\bibitem{Landry_Hua_2018}
Guillaume Landry and Chia‐ho Hua.
\newblock Current state and future applications of radiological image guidance for particle therapy.
\newblock {\em Medical Physics}, 45(11), November 2018.

\bibitem{Rabe2025283}
Moritz Rabe, Christopher Kurz, Adrian Thummerer, and Guillaume Landry.
\newblock Artificial intelligence for treatment delivery: image-guided radiotherapy.
\newblock {\em Strahlentherapie und Onkologie}, 201(3):283 – 297, 2025.

\bibitem{RIOU2024603}
Olivier Riou, Jessica Prunaretty, and Morgan Michalet.
\newblock Personalizing radiotherapy with adaptive radiotherapy: Interest and challenges.
\newblock {\em Cancer/Radiothérapie}, 28(6):603--609, 2024.
\newblock 35e congrès de la Société française de radiothérapie oncologique.

\bibitem{Mastella2025}
Edoardo Mastella, Francesca Calderoni, Luigi Manco, Martina Ferioli, Serena Medoro, Alessandro Turra, Melchiore Giganti, and Antonio Stefanelli.
\newblock A systematic review of the role of artificial intelligence in automating computed tomography-based adaptive radiotherapy for head and neck cancer.
\newblock {\em Physics and Imaging in Radiation Oncology}, 33:100731, January 2025.

\bibitem{HUIJBEN2024}
Evi~M.C. Huijben, Maarten~L. Terpstra, Arthur~Jr. Galapon, Suraj Pai, Adrian Thummerer, Peter Koopmans, Manya Afonso, Maureen {van Eijnatten}, Oliver Gurney-Champion, Zeli Chen, Yiwen Zhang, Kaiyi Zheng, Chuanpu Li, Haowen Pang, Chuyang Ye, Runqi Wang, Tao Song, Fuxin Fan, Jingna Qiu, Yixing Huang, Juhyung Ha, Jong {Sung Park}, Alexandra Alain-Beaudoin, Silvain Bériault, Pengxin Yu, Hongbin Guo, Zhanyao Huang, Gengwan Li, Xueru Zhang, Yubo Fan, Han Liu, Bowen Xin, Aaron Nicolson, Lujia Zhong, Zhiwei Deng, Gustav Müller-Franzes, Firas Khader, Xia Li, Ye~Zhang, Cédric Hémon, Valentin Boussot, Zhihao Zhang, Long Wang, Lu~Bai, Shaobin Wang, Derk Mus, Bram Kooiman, Chelsea~A.H. Sargeant, Edward~G.A. Henderson, Satoshi Kondo, Satoshi Kasai, Reza Karimzadeh, Bulat Ibragimov, Thomas Helfer, Jessica Dafflon, Zijie Chen, Enpei Wang, Zoltan Perko, and Matteo Maspero.
\newblock Generating synthetic computed tomography for radiotherapy: {SynthRAD2023} challenge report.
\newblock {\em Medical Image Analysis}, 97:103276, 2024.

\bibitem{Raggio2025}
Ciro~Benito Raggio, Mathias~Krohmer Zabaleta, Nils Skupien, Oliver Blanck, Francesco Cicone, Giuseppe~Lucio Cascini, Paolo Zaffino, Lucia Migliorelli, and Maria~Francesca Spadea.
\newblock {FedSynthCT-Brain: A} federated learning framework for multi-institutional brain mri-to-ct synthesis.
\newblock {\em Computers in Biology and Medicine}, 192:110160, 2025.

\bibitem{kaissis2020secure}
Georgios~A Kaissis, Marcus~R Makowski, Daniel R{\"u}ckert, and Rickmer~F Braren.
\newblock Secure, privacy-preserving and federated machine learning in medical imaging.
\newblock {\em Nature Machine Intelligence}, 2(6):305--311, 2020.

\bibitem{guan2024federated}
Hao Guan, Pew-Thian Yap, Andrea Bozoki, and Mingxia Liu.
\newblock Federated learning for medical image analysis: A survey.
\newblock {\em Pattern Recognition}, 151:110424, 2024.

\bibitem{Sandhu2023}
Sukhveer~Singh Sandhu, Hamed~Taheri Gorji, Pantea Tavakolian, Kouhyar Tavakolian, and Alireza Akhbardeh.
\newblock Medical imaging applications of federated learning.
\newblock {\em Diagnostics}, 13(19):3140, October 2023.

\bibitem{DALMAZ2024103121}
Onat Dalmaz, Muhammad~U. Mirza, Gokberk Elmas, Muzaffer Ozbey, Salman~U.H. Dar, Emir Ceyani, Kader~K. Oguz, Salman Avestimehr, and Tolga Çukur.
\newblock One model to unite them all: Personalized federated learning of multi-contrast mri synthesis.
\newblock {\em Medical Image Analysis}, 94:103121, 2024.

\bibitem{Hernandez-Cruz2024}
Netzahualcoyotl Hernandez-Cruz, Pramit Saha, Md~Mostafa~Kamal Sarker, and J.~Alison Noble.
\newblock {Review of Federated Learning and Machine Learning-Based Methods for Medical Image Analysis}.
\newblock {\em Big Data and Cognitive Computing}, 8(9):99, August 2024.

\bibitem{thummerer2025synthrad2025}
Adrian Thummerer, Erik van~der Bijl, Arthur~Jr Galapon, Florian Kamp, Mark Savenije, Christina Muijs, Shafak Aluwini, Roel J. H.~M. Steenbakkers, Stephanie Beuel, Martijn P.~W. Intven, Johannes~A. Langendijk, Stefan Both, Stefanie Corradini, Viktor Rogowski, Maarten Terpstra, Niklas Wahl, Christopher Kurz, Guillaume Landry, and Matteo Maspero.
\newblock {SynthRAD2025 Grand Challenge} dataset: generating synthetic {CTs} for radiotherapy, 2025.

\bibitem{Rusanov2022}
Branimir Rusanov, Ghulam~Mubashar Hassan, Mark Reynolds, Mahsheed Sabet, Jake Kendrick, Pejman Rowshanfarzad, and Martin Ebert.
\newblock Deep learning methods for enhancing cone-beam {CT} image quality toward adaptive radiation therapy: A systematic review.
\newblock {\em Medical Physics}, 49(9):6019--6054, 2022.

\bibitem{Ronneberger2015}
Olaf Ronneberger, Philipp Fischer, and Thomas Brox.
\newblock U-net: Convolutional networks for biomedical image segmentation.
\newblock In Nassir Navab, Joachim Hornegger, William~M. Wells, and Alejandro~F. Frangi, editors, {\em Medical Image Computing and Computer-Assisted Intervention -- MICCAI 2015}, pages 234--241, Cham, 2015. Springer International Publishing.

\bibitem{isola2018}
Phillip Isola, Jun-Yan Zhu, Tinghui Zhou, and Alexei~A. Efros.
\newblock Image-to-image translation with conditional adversarial networks, 2018.

\bibitem{Chen2022}
Xinyuan Chen, Yuxiang Liu, Bining Yang, Ji~Zhu, Siqi Yuan, Xuejie Xie, Yueping Liu, Jianrong Dai, and Kuo Men.
\newblock A more effective {CT} synthesizer using transformers for cone-beam {CT-guided} adaptive radiotherapy.
\newblock {\em Frontiers in Oncology}, Volume 12 - 2022, 2022.

\bibitem{LaGrecaSaintEsteven2023}
Agustina {La Greca Saint-Esteven}, Ricardo {Dal Bello}, Mariia Lapaeva, Lisa Fankhauser, Bertrand Pouymayou, Ender Konukoglu, Nicolaus Andratschke, Panagiotis Balermpas, Matthias Guckenberger, and Stephanie Tanadini-Lang.
\newblock Synthetic computed tomography for low-field magnetic resonance-only radiotherapy in head-and-neck cancer using residual vision transformers.
\newblock {\em Physics and Imaging in Radiation Oncology}, 27:100471, 2023.

\bibitem{Sheller2019-jo}
Micah~J Sheller, G~Anthony Reina, Brandon Edwards, Jason Martin, and Spyridon Bakas.
\newblock {Multi-Institutional} deep learning modeling without sharing patient data: A feasibility study on brain tumor segmentation.
\newblock {\em Brainlesion}, 11383:92--104, January 2019.

\bibitem{Wenqi2019}
Wenqi Li, Fausto Milletar{\`i}, Daguang Xu, Nicola Rieke, Jonny Hancox, Wentao Zhu, Maximilian Baust, Yan Cheng, S{\'e}bastien Ourselin, M.~Jorge Cardoso, and Andrew Feng.
\newblock Privacy-preserving federated brain tumour segmentation.
\newblock In Heung-Il Suk, Mingxia Liu, Pingkun Yan, and Chunfeng Lian, editors, {\em Machine Learning in Medical Imaging}, pages 133--141, Cham, 2019. Springer International Publishing.

\bibitem{Daiqing_Li2020}
Daiqing Li, Amlan Kar, Nishant Ravikumar, Alejandro~F Frangi, and Sanja Fidler.
\newblock {Fed-Sim}: Federated simulation for medical imaging, 2020.

\bibitem{Chang2020}
Qi~Chang, Hui Qu, Yikai Zhang, Mert Sabuncu, Chao Chen, Tong Zhang, and Dimitris Metaxas.
\newblock Synthetic learning: Learn from distributed asynchronized discriminator gan without sharing medical image data, 2020.

\bibitem{McMahan2017}
Brendan McMahan, Eider Moore, Daniel Ramage, Seth Hampson, and Blaise Aguera~y Arcas.
\newblock {Communication-Efficient Learning of Deep Networks from Decentralized Data}.
\newblock In Aarti Singh and Jerry Zhu, editors, {\em Proceedings of the 20th International Conference on Artificial Intelligence and Statistics}, volume~54 of {\em Proceedings of Machine Learning Research}, pages 1273--1282. PMLR, 20--22 Apr 2017.

\bibitem{Wang2023}
Jinbao Wang, Guoyang Xie, Yawen Huang, Jiayi Lyu, Feng Zheng, Yefeng Zheng, and Yaochu Jin.
\newblock {FedMed-GAN}: Federated domain translation on unsupervised cross-modality brain image synthesis.
\newblock {\em Neurocomputing}, 546:126282, 2023.

\bibitem{SynthRAD2025DatasetV2}
Adrian Thummerer, Erik van~der Bijl, Arthur~Jr. Galapon, Florian Kamp, and Matteo Maspero.
\newblock {SynthRAD2025 Grand Challenge} dataset: Training, 2025.

\bibitem{beutel2022flowerfriendlyfederatedlearning}
Daniel~J. Beutel, Taner Topal, Akhil Mathur, Xinchi Qiu, Javier Fernandez-Marques, Yan Gao, Lorenzo Sani, Kwing~Hei Li, Titouan Parcollet, Pedro Porto~Buarque de~Gusmão, and Nicholas~D. Lane.
\newblock {Flower: A Friendly Federated Learning Research Framework}, 2022.

\bibitem{pytorch}
Adam Paszke, Sam Gross, Francisco Massa, Adam Lerer, James Bradbury, Gregory Chanan, Trevor Killeen, Zeming Lin, Natalia Gimelshein, Luca Antiga, Alban Desmaison, Andreas K\"{o}pf, Edward Yang, Zach DeVito, Martin Raison, Alykhan Tejani, Sasank Chilamkurthy, Benoit Steiner, Lu~Fang, Junjie Bai, and Soumith Chintala.
\newblock {\em PyTorch: an imperative style, high-performance deep learning library}.
\newblock Curran Associates Inc., 2019.

\bibitem{cardoso2022monaiopensourceframeworkdeep}
M.~Jorge Cardoso, Wenqi Li, Richard Brown, Nic Ma, Eric Kerfoot, Yiheng Wang, Benjamin Murrey, Andriy Myronenko, Can Zhao, Dong Yang, Vishwesh Nath, Yufan He, Ziyue Xu, Ali Hatamizadeh, Andriy Myronenko, Wentao Zhu, Yun Liu, Mingxin Zheng, Yucheng Tang, Isaac Yang, Michael Zephyr, Behrooz Hashemian, Sachidanand Alle, Mohammad~Zalbagi Darestani, Charlie Budd, Marc Modat, Tom Vercauteren, Guotai Wang, Yiwen Li, Yipeng Hu, Yunguan Fu, Benjamin Gorman, Hans Johnson, Brad Genereaux, Barbaros~S. Erdal, Vikash Gupta, Andres Diaz-Pinto, Andre Dourson, Lena Maier-Hein, Paul~F. Jaeger, Michael Baumgartner, Jayashree Kalpathy-Cramer, Mona Flores, Justin Kirby, Lee A.~D. Cooper, Holger~R. Roth, Daguang Xu, David Bericat, Ralf Floca, S.~Kevin Zhou, Haris Shuaib, Keyvan Farahani, Klaus~H. Maier-Hein, Stephen Aylward, Prerna Dogra, Sebastien Ourselin, and Andrew Feng.
\newblock {MONAI}: An open-source framework for deep learning in healthcare, 2022.

\bibitem{Vicario2022}
Celia~Martín Vicario, Florian Kordon, Felix Denzinger, Jan Siad~El Barbari, Maxim Privalov, Jochen Franke, Andreas Maier, and Holger Kunze.
\newblock Normalization techniques for {CNN based analysis} of surgical cone beam ct volumes.
\newblock {\em Medical Imaging 2022: Image Processing}, page~85, March 2022.

\bibitem{li2020federatedepochsopt}
Tian Li, Anit~Kumar Sahu, Manzil Zaheer, Maziar Sanjabi, Ameet Talwalkar, and Virginia Smith.
\newblock Federated optimization in heterogeneous networks, 2020.

\bibitem{Kumar2024}
Vikas Kumar, Manoj Sharma, R.~Jehadeesan, B.~Venkatraman, and Debdoot Sheet.
\newblock Simulating cross-modal medical images using multi-task adversarial learning of a deep convolutional neural network.
\newblock {\em International Journal of Imaging Systems and Technology}, 34(4):e23113, 2024.

\end{thebibliography}

\end{document}